\documentclass[12pt]{article}
\usepackage{amsmath}
\usepackage{setspace}
\usepackage{subfig}
\usepackage{graphicx}
\usepackage{textcomp}
\usepackage{subfig}
\usepackage{array, adjustbox}
\usepackage{authblk}
\usepackage[top=0.7in, left=0.6in, right=0.6in, bottom=0.7in]{geometry}
\usepackage{lineno}
\providecommand{\keywords}[1]{\textbf{Keywords---} #1}

\usepackage{cite}

\title{\textbf{Characterization of Vegetation and Soil Scattering Mechanisms across Different Biomes using P-band SAR Polarimetry}}

\author[1,2]{Seyed Hamed Alemohammad\thanks{Corresponding author: S. H. Alemohammad (hamed\_al@mit.edu).}}
\author[3]{Alexandra G. Konings}
\author[4]{Thomas Jagdhuber}
\author[5]{Mahta Moghaddam}
\author[1]{Dara Entekhabi}
\affil[1]{\small{Department of Civil and Environmental Engineering, Massachusetts Institute of Technology}}
\affil[2]{Department of Earth and Environmental Engineering, Columbia University}
\affil[3]{Department of Earth System Science, Stanford University}
\affil[4]{Microwaves and Radar Institute, German Aerospace Center (DLR)}
\affil[5]{Department of Electrical Engineering, University of Southern California}

\date{}
\normalsize

\begin{document}
\onehalfspacing
\maketitle
\doublespacing
\begin{abstract}
	Understanding the scattering mechanisms from the ground surface in the presence of different vegetation densities is necessary for the interpretation of P-band Synthetic Aperture Radar (SAR) observations and for the design of geophysical retrieval algorithms. In this study, a quantitative analysis of vegetation and soil scattering mechanisms estimated  from the observations  of  an airborne P-band SAR  instrument  across  nine  different  biomes  in North  America is presented.  The goal is to apply a hybrid (model- and eigen- based) three component decomposition approach to separate the contributions of surface, double-bounce and vegetation volume scattering across a wide range of biome conditions. The decomposition makes no prior assumptions about vegetation structure. We characterize the dynamics of the decomposition across different North American biomes and assess their characteristic range. Impacts of vegetation cover seasonality and soil surface roughness on the contributions of each scattering mechanism are also investigated. Observations used here are part of the NASA Airborne Microwave Observatory of Subcanopy and Subsurface (AirMOSS) mission and data have been collected between 2013 and 2015.  
\end{abstract}
\keywords{AirMOSS, P-Band, Polarimetric Decomposition, SAR, Soil Moisture, Vegetation}
\section{Introduction}

Soil moisture is the key state variable that controls the terrestrial water, carbon, and energy fluxes between land surface and atmospheric boundary layer, mainly by regulating photosynthesis and surface evaporation \cite{Seneviratne2010}. By constraining the partitioning of energy fluxes between latent and sensible heat fluxes in water-limited regions, soil moisture also plays a significant role in the prediction skill of weather and climate models \cite{Entekhabi1996, Koster2010}. Soil moisture has a memory that captures the anomalies in precipitation and radiation and can be used to identify regions of strong feedback between land surface and atmospheric boundary layer \cite{McColl2017}. Moreover, vegetation stress, and subsequently photosynthetic activity, depends on the amount of water available through the roots (as well as atmospheric conditions). Therefore, knowledge of Root Zone Soil Moisture (RZSM) (and where applicable, interaction of roots, soil moisture, and the water table) are necessary to accurately model evapotranspiration seasonality \cite{Thompson2011}. Characterization of the spatio-temporal patterns of soil moisture with depth, therefore, enables improved predictions of the response of plants to the changing climate. Thus, remotely sensed, large-scale estimates of root-zone soil moisture have a number of operational and scientific use in hydrometeorology and ecology, if they can be obtained. 

The penetration depth associated with microwave remote sensing soil moisture increases as the electromagnetic frequency of the measurement decreases \cite{Ulaby2014}. Current global microwave satellite observations of soil moisture are limited to those at L-band frequency and higher due to spectrum availability, readiness of science algorithms and technological restrictions of obtaining reasonable spatio-temporal resolution from low-earth orbit satellites \cite{Entekhabi2014, Kerr2010}. However, L-band instruments, are sensitive to only a few centimeters of the top soil layer. Detection of RZSM requires P-band instruments, which have a penetration depth of several tens of centimeters, depending on soil texture, the profile of soil moisture content and vegetation cover \cite{Moghaddam2007, Konings2014}. The future BIOMASS mission will carry a fully polarimetric P-band Synthetic Aperture Radar (SAR) instrument and provide an opportunity to estimate RZSM globally (except over North America and Europe) from satellite-based SAR observations \cite{Toan2011, Carreiras2017}. 

The Airborne Microwave Observatory of Subcanopy and Subsurface (AirMOSS) mission (a NASA Earth Venture-1 project) is the first P-band airborne campaign to estimate RZSM and use that to better characterize Net Ecosystem Exchange (NEE) across North America \cite{Allen2010}. AirMOSS employs an airborne fully polarimetric SAR operating at P-band (430 MHz) to monitor the dynamics of RZSM across ten sites in North America, covering nine different biomes representative of the entire North American continent (Table \ref{campaigns}). The biomes across AirMOSS sites range from tropical and temperate forest to boreal transitional forest and evergreen needle-leaf as well as cropland, woody savanna, grassland and shrubland. AirMOSS backscatter estimates have a high radiometric calibration accuracy of 0.5 dB cite{Chapin2015}, and a noise equivalent $\sigma^\circ$ of -40 dB \cite{Tabatabaeenejad2015} which makes them suitable for accurate soil and vegetation parameter estimation (although the data require a phase calibration for polarimetric applications, as further discussed below).

AirMOSS successfully conducted four years (2012-2015) of airborne campaigns, with 186 science flights each covering an area of approximately 100 km $\times$ 25 km. Each site was visited for 2 or 3 campaigns per year. For each campaign, 2 to 3 overflights were performed a few days apart. Data from these campaigns provide an opportunity to develop and validate new algorithms for RZSM, vegetation and surface properties retrieval and to analyze and quantify the contributions of volume and surface scattering, and their interactions to P-band signals across a wide range of vegetation types. In this study, we provide a quantitative analysis of the variability of surface and vegetation parameters across the campaign sites.  The AirMOSS radar calibration model was updated in April 2013, so we restrict our study to observations from the period April 2013 through September 2015 to ensure a consistent calibration accuracy/quality.

\begin{table}
	\renewcommand{\arraystretch}{1.5}
	\caption{List of AirMOSS campaign sites and their biome type}
	\label{campaigns}
	\centering
	\begin{adjustbox}{width=\textwidth}
		\begin{tabular}{|c|c|}
			\hline
			\textbf{Name \& Location} & \textbf{Biome Type } \\
			\hline
			\hline
			BERMS, Saskatchewan, Canada & Boreal forest / evergreen needle-leaf, mixed forest, cropland \\
			\hline
			Howland  Forest, ME, USA & Boreal transitional / mixed forest \\
			\hline
			Harvard Forest, MA, USA & Boreal transitional / mixed forest \\
			\hline
			Duke Forest, NC, USA & Temperate forest / mixed forest, cropland \\
			\hline
			Metolius, OR, USA & Temperate forest / evergreen needle-leaf \\
			\hline
			MOISST, Marena, OK, USA & Temperate grasslands / crops \\
			\hline
			Tonzi Ranch, CA, USA & Mediterranean forest / woody savanna \\
			\hline
			Walnut Gulch, AZ, USA & Desert and shrub / open shrubland and grassland \\
			\hline
			Chamela, Mexico & Subtropical dry forest / broadleaf deciduous, crops, woody savanna \\
			\hline
			La Selva, Costa Rica & Tropical moist forest / evergreen broadleaf, crops \\ 
			\hline	
		\end{tabular}
	\end{adjustbox}
\end{table}

Unlike at L-band, the variability of the soil moisture with depth is significant over P-band penetration ranges and must be accounted for, which means calculating only a single equivalent soil moisture value may result in values that do not represent the average profile due to the effects of subsurface reflections \cite{Konings2014}. Hence, two different approaches have previously been used for retrieval of soil moisture profiles from AirMOSS observations using modeled vegetation and ancillary vegetation parameters. For campaign sites with mono-species woody or non-woody vegetation (including mono-species forested sites), the approach is to model the vegetation using a detailed scattering model that consists of a stem layer and a canopy layer \cite{Burgin2011}, and assume a second order polynomial shape for the soil moisture profile. Parameters of the vegetation volume scattering model are derived based on field measurements, and the coefficients of the polynomial profile are retrieved using snapshot measurements \cite{Tabatabaeenejad2015}. This approach also assumes a temporally and spatially constant value for the surface roughness parameter, determined by site-specific calibration, for each land cover type to reduce the number of unknowns in the retrieval. 

A second retrieval algorithm using AirMOSS observations is focused on campaign sites with multi-species woody vegetation cover \cite{Truong-Loi2015}. This approach uses a semiempirical inversion model to estimate soil moisture profile, surface roughness and the aboveground biomass. The parameters of the scattering model are estimated by regressing the semiempirical model to forward estimates of the full scattering model using field measurements from the Forest Inventory Analysis \cite{FIA2013}. In order to increase the number of observations and make the system of equations well-defined, a time series scheme is designed that assumes constant surface roughness and above ground biomass across the three flight days of each AirMOSS campaign (2 to 3 observations over 7-10 days). RZSM retrievals from both of these algorithms meet the AirMOSS mission requirements of an unbiased Root Mean Squared Error (ubRMSE) of 0.05 $m^3/m^3$ for soil moisture over the AirMOSS sites when validated against ground measurements \cite{Tabatabaeenejad2015, Truong-Loi2015}.

Using vegetation (and roughness) parameters that are dependent on site-specific field measurements limits the applicability of these methods outside of the United States (where FIA data are not available) and in areas where \textit{in situ} measurements may be logistically difficult and expensive. Furthermore, these approaches limit the applicability of the retrieval algorithm across diverse and large scale land covers and may create errors due to variability in vegetation structure even across a single site. Hence, the transferability of the approach to other test sites or remote regions is hardly given. However, the design of retrieval algorithms that require fewer parameters is complicated by the limited amount of information contained in the backscattering coefficients. Use of the coherent fully polarimetric observations, which was not implemented in the existing AirMOSS retrieval algorithms, provides a means to overcome this potential problem.

A quantitative understanding of scattering mechanisms from land surfaces with vegetation cover is required for interpreting P-band SAR observations. Understanding the relative role of attenuated ground backscatter, vegetation volume backscatter and interactions term based on the SAR observations serves as a guide to the design of parsimonious RZSM retrieval algorithms. The goal of this study is to quantify the contributions of ground (surface and double-bounce) and volume scattering across the wide range of vegetation covers in fully polarimetric P-band SAR observations. Such an extensive analysis was not conducted before due to the lack of fully polarimetric P-band data with reasonable SNR over different land covers/biomes and across different seasons. 

We use AirMOSS coherent fully polarimetric observations (phase and amplitude information). To begin, it is necessary to overcome the lack of a dedicated phase calibration in the AirMOSS observations. First, we preprocess the observations by merging the four flight lines in each campaign, removing pixels with high topographic slope and calibrating the polarimetric phase of the observations. Then, we apply a fully polarimetric decomposition model to estimate the contribution of each scattering mechanism to the total backscattering power. Vegetation scattering is modeled using a cloud of randomly-oriented dipoles, without the need for prior assumptions on vegetation parameters. Applying the estimation approach to observations across all AirMOSS campaign sites, we characterize the temporal and spatial differences in the relative contributions of the different scattering mechanisms. These results can provide guidance for the design of future low-frequency RZSM retrieval algorithm. 

The rest of the study is organized as following: Section \ref{phase_cal} reviews the data pre-processing and presents estimates of phase bias from AirMOSS observations. Section \ref{model} describes the fully polarimetric model used to decompose the scattering mechanisms, and Section \ref{estimation} outlines the estimation steps. Section \ref{results} presents the results, and conclusions are provided in Section \ref{conclusion}.

\section{Data Preprocessing}
\label{phase_cal}

The AirMOSS instrument has three independent polarization channels (i.e. HH, VV, VH) measured with amplitude and phase. The incidence angle of the processed image products ranges from 25$^\circ$ to 55$^\circ$ \cite{Chapin2012}. In our decomposition model, we assume the observations follow reflection symmetry (Section \ref{model}). AirMOSS measurements in each campaign site include four adjacent flight lines with few overlapping pixels. For the purpose of this study, we have merged the flight lines for each campaign and for the overlapping pixels, measurements with an incidence angle closer to $40^\circ$ are selected (chosen to be in the middle range of the incidence angles from AirMOSS instrument). Moreover, pixels with high topographic slope are excluded from this analysis. Steep surfaces act like a rotated surface, and the assumption of reflection symmetry is violated \cite{Freeman1998}. Slopes are calculated using digital elevation data and provided by the science team as part of the ancillary data in each campaign. We choose a threshold of 15\% slope and removed pixels that have a higher slope in either North-South or East-West direction. This results in removal of between 5\% to 45\% of the pixels across different sites. 

The use of coherent polarimetric information from a SAR instrument requires the measurements to be phase calibrated. Airborne SAR instruments are often calibrated using corner reflectors \cite{VanZyl1990}, but no corner reflectors were placed across any of the campaign sites. Therefore, AirMOSS polarimetric measurements are not phase calibrated.

Here, we use the distribution of measurements across low vegetation pixels (as a canonical target) for phase calibration (similar to the approach developed by \cite{Zebker1990}). Unlike the approach proposed by \cite{Zebker1990}, we use cross-pol observations to identify low vegetation pixels. For these pixels, the phase of the $S_{HH}S_{VV}^*$ in backscattering should ideally be zero \cite{Guissard1994}. Therefore, we use the phase distribution of this signal across many bare soil / low vegetation pixels to estimate the phase offset from zero, if any. 

Static land cover maps cannot account for seasonally varying changes in whether a pixel has low vegetation cover, and underestimate the number of low-vegetation pixels. Instead, we use the data-driven criteria of $-40 [dB] < S_{HV}S_{HV}^* < -25 [dB]$ to identify bare soil and low vegetation pixels. This criterion is selected based on the fact that vegetation cover leads to relatively high cross-polarized backscattering; therefore, a low cross-polarized backscatter value indicates low or no vegetation coverage. Soil roughness can also contribute to the cross-polarized signal, but we expect that to be negligible for P-band and non-tilled soils. To remove observations that are below the noise level, the lower bound for this criteria is set to the noise equivalent $\sigma^\circ$ of AirMOSS instrument \cite{Chapin2012, Hensley2015}. To increase the accuracy of this data-driven detection, we further excluded pixels that are classified as either developed, open water or wetland based on the NLCD land cover data. 

\begin{figure}
	\centering
	\includegraphics[width=0.7\linewidth]{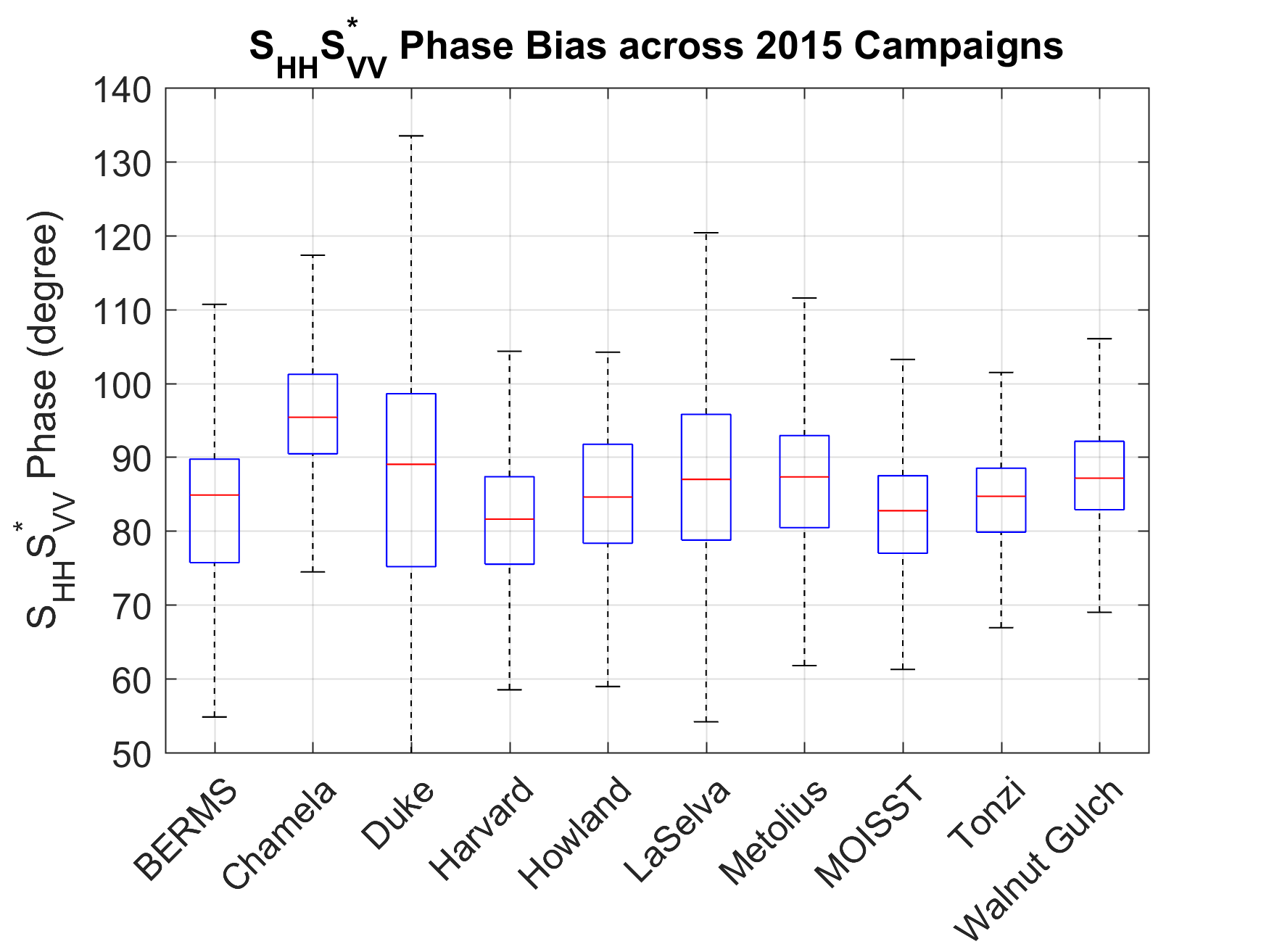}
	\caption{Distribution of $S_{HH}S_{VV}^*$ across different sites for one campaign in 2015. The phase data from pixels that have a $|S_{HV}|^2$ in the range [-32 -26] dB are included in this figure. The central red line indicates the median, the edges of the box are 25th and 75th percentiles, the whiskers show the most extreme values.}
	\label{phase_dist}
\end{figure}

We applied this approach to all the campaigns used in this study. Figure \ref{phase_dist} shows the distribution of $S_{HH}S_{VV}^*$ phases across bare soil / low vegetation pixels for one of the campaigns at each site conducted in 2015 (as an example of the variability of the phase bias across different campaign/sites). We use the median of phase across these pixels in each flight campaign as the bias in the phase information, and correct $S_{HH}S_{VV}^*$ observation for that flight by removing the phase bias. Therefore, the phase bias varied from each campaign and site to another. For the 2015 flights shown Figure \ref{phase_dist}, the estimated bias ranges from 82$^\circ$ to 95$^\circ$.

\section{Forward Physical Model of Vegetated Surface}
\label{model}

A key challenge in monitoring soil moisture using radar measurements is that the interaction of signal with vegetation structure and water content is complex for a general vegetation cover. Moreover, the backscatter signal has contributions directly from the vegetation and bounced signal between the surface and vegetation (double-bounce) together with the backscatter from the surface. Since the vegetation characteristics are not known a priori, they add more unknowns to the moisture estimation problem. This problem is exacerbated when using low-frequency (e.g., P-band) observations to estimate RZSM. Soil moisture can vary significantly across the root-zone, with the shape of the profile varying in time. Interactions and phase delays between reflections from different soil layers cause the equivalent half-space soil moisture values to differ significantly from the true average soil moisture profile, so that soil moisture at multiple depths has to be retrieved simultaneously \cite{Konings2014}.

Two main approaches have been developed to decompose fully polarimetric SAR observations. First, eigen-based decomposition that is a mathematical data-driven technique to interpret eigenvalues and eigenvectors of the coherency matrix as physical mechanisms contributing to the backscattering observation \cite{Holm1988, VanZyl1989, Vanzyl1992, Cloude1996}. Second, model-based decomposition that develops a physical scattering model and retrieves the parameters using the covariance or coherency matrix of observations \cite{Freeman1993, Freeman1998, Freeman2007, Yamaguchi2005}. While, in general, it is not straightforward to provide physical interpretation of eigenvalue-based decompositions, model-based approaches also suffer from large number of unknowns that limit their applicability. Nevertheless, both approaches are widely used in decomposing SAR observations and retrieving soil moisture and vegetation parameters \cite{Moghaddam2000, Hajnsek2009, Kim2009, VanZyl2011, Arii2011, Jagdhuber2012b, Jagdhuber2013, Kim2014, Jagdhuber2015, Alemohammad2016, He2016}. Here, we use a novel comprise and combine the model-based decomposition approach of \cite{Freeman1998} with an eigen-based alpha-angle approach (as in \cite{Jagdhuber2015}) to benefit from the capabilities of both techniques. 

We use a three-component scattering model and decompose the total backscattering signal to contributions from ground surface scattering, vegetation volume scattering and double-bounce scattering between the vegetation and ground surface \cite{Freeman1993, Freeman1998, Lee2009, Yamaguchi2006}:
\begin{equation}
	\langle T\rangle^{obs} = T^S + T^V + T^D
	\label{}
\end{equation}

in which $S$, $V$, and $D$ denote ground surface, volume, and double-bounce scattering mechanisms, respectively. $\langle \rangle$ denotes spatial ensemble average in the data processing. $T$ represents the scattering coherency matrix:
\begin{eqnarray}
	\nonumber T && = 
	\begin{bmatrix}
		T_{11} & T_{12} & T_{13}\\
		T^*_{12} & T_{22} & T_{23}\\
		T^*_{13} & T^*_{23} & T_{33}\\
	\end{bmatrix}\\
	&& =\frac{1}{2}
	\begin{bmatrix}
		\langle|S_{HH} + S_{VV}|^2\rangle & \langle (S_{HH} + S_{VV})(S_{HH} - S_{VV})^*\rangle & 2\langle (S_{HH} + S_{VV})S^*_{HV}\rangle\\
		\langle (S_{HH} - S_{VV})(S_{HH} + S_{VV})^*\rangle	   & \langle|S_{HH} - S_{VV}|^2\rangle & 2\langle (S_{HH} - S_{VV})S^*_{HV}\rangle\\
		2\langle S_{HV}(S_{HH} + S_{VV})^*\rangle & 2\langle S_{HV}(S_{HH} - S_{VV})^*\rangle & 4\langle|S_{HV}|^2\rangle\\
	\end{bmatrix}
\end{eqnarray}

where superscript $^*$ denotes complex conjugate. For the model used in this study, we assume reflection symmetry (which is validated by the observations, not shown here); therefore, in the $T^{model}$, $ T_{13} = T_{23} = 0$.

Vegetation is modeled assuming the vegetation layer consists of cylindrical particles that are randomly oriented with a uniform probability density function (PDF) [8, 9]. The coherency matrix for vegetation volume scattering is \cite{Yamaguchi2005, Yamaguchi2006}:
\begin{equation}
	T^V = f_v
	\begin{bmatrix}
		0.50 & 0 & 0\\
		0 & 0.25 & 0\\
		0 & 0 & 0.25
	\end{bmatrix}
	\label{(T_V)}
\end{equation}

in which $f_v$ represents vegetation volume scattering intensity. For surface and double-bounce scattering we use the $\alpha$ scattering definition to find their corresponding contribution \cite{Cloude2010}. The $\alpha$ scattering angle is calculated using an eigen-based decomposition of the coherency matrix, and defined as the inverse cosine of the length of the first element of the first eigenvector of the coherency matrix $T^{rem}$. A detailed derivation of the $\alpha$ scattering angle is presented in \cite{Cloude1997, Cloude2001, Jagdhuber2012}. Assuming surface and double-bounce scattering components are orthogonal rank-1, the coherency matrix for direct backscattering from the surface would be \cite{Jagdhuber2015, Cloude2010}:
\begin{equation}
	T^S = f_s
	\begin{bmatrix}
		cos^2\alpha_s & -sin\alpha_s & 0\\
		-cos\alpha_s & sin^2\alpha_s & 0\\
		0 & 0 & 0
	\end{bmatrix}
	\label{(T_S)}
\end{equation}

in which $f_s$ represents intensity of surface scattering. Double-bounce scattering is represented with \cite{Jagdhuber2015, Cloude2010}:
\begin{equation}
	T^D = f_d
	\begin{bmatrix}
		sin^2\alpha_d & cos\alpha_d & 0\\
		sin\alpha_d & cos^2\alpha_d & 0\\
		0 & 0 & 0
	\end{bmatrix}
	\label{(T_D)}
\end{equation}

in which $f_d$ represents the intensity of double-bounce scattering.  

The model presented here can capture the interactions of a low-frequency SAR signal within the vegetated surface with relatively simple representation of volume scattering. The focus of this study is on characterizing the contribution of each scattering mechanism; therefore, we estimate $\alpha_s$ and $\alpha_d$ parameters together with $f_s$, $f_d$, and $f_v$ rather than the explicit surface and vegetation reflectivities (or dielectric properties) and surface roughness. This analysis is aimed at estimating the three scattering decomposition across a wide range of vegetation covers (biomes) using P-band polarimetric observations alone (no ancillary information is used).

\section{Estimation Approach}
\label{estimation}

In this section, we outline the estimation approach to quantify the contribution of each scattering mechanism. We use a hybrid (model- and eigen- based) approach and use the physical model introduced in the previous section along with an orthogonality criterion on the two eigen-based $\alpha$ scattering angles to separate the surface and double-bounce scattering contributions \cite{Cloude2010, Jagdhuber2015}.

Observations from AirMOSS instrument are provided in terms of the elements of the covariance matrix. Therefore, we use the special unitary transformation matrix to transform the covariance matrix to coherency matrix \cite{Lee2009, Cloude2010}. The observed coherency matrix after implementing the phase calibration is:
\begin{equation}
	T^{obs} = \frac{1}{2}
	\begin{bmatrix}
		\langle|S^{obs}_{HH} + S^{obs}_{VV}|^2\rangle & \langle (S^{obs}_{HH} + S^{obs}_{VV})(S^{obs}_{HH} - S^{obs}_{VV})^*\rangle & 0\\
		\langle (S^{obs}_{HH} - S^{obs}_{VV})(S^{obs}_{HH} + S^{obs}_{VV})^*\rangle	   & \langle|S^{obs}_{HH} - S^{obs}_{VV}|^2\rangle & 0\\
		0 & 0 & 4\langle|S^{obs}_{HV}|^2\rangle\\
	\end{bmatrix}
\end{equation}

The first step in the estimation process is to estimate $f_v$ and remove the vegetation contribution from the observations. Based on the model presented in the previous section, only volume scattering contributes to the cross-pol channel; therefore, we can estimate $f_v$ directly from the cross-pol observation ($T_{33}$). Previous studies have shown that the cross-polarized channel can have contributions from other scattering mechanisms (ground surface and double-bounce) and this might result in a higher value for $f_v$ that is physically impossible \cite{VanZyl2011}. Therefore, $f_v$ estimation should be bounded so that after removing the vegetation contribution the remained power in surface and double-bounce scattering is non-negative (a mathematical constraint). This implies that the eigenvalues of the $T^{rem}$ coherency matrix should be non-negative \cite{VanZyl2011}:
\begin{equation}
	T^{rem} = T^{obs} - f_v
	\begin{bmatrix}
		0.50 & 0 & 0\\
		0 & 0.25 & 0\\
		0 & 0 & 0.25
	\end{bmatrix}
\end{equation}

The objective function to estimate $f_v$ by incorporating the upper bound limit is:
\begin{eqnarray}
	\nonumber	f_{v} &&= \max(f_{v})\\
	\text{subject to} && \text{eig}(T^{rem}) \geq 0	
\end{eqnarray}  

Here, the eigenvalues of $T^{rem}$ are numerically calculated for different values of $f_v$, and the maximum physically feasible $f_v$ is estimated. 

Next, we estimate $\alpha_s$, $\alpha_d$, $f_s$ and $f_d$ using the equations provided in \cite{Jagdhuber2015}. These equations provide solutions for these four variables with an ambiguity for $\alpha_s$ and $\alpha_d$ which is resolved based on an orthogonality condition ($\alpha_s=\frac{\pi}{2} - \alpha_d$), inherent in the physics of the alpha scattering model \cite{Cloude2001}. Among the two estimated $\alpha's$, the one that is smaller than $\frac{\pi}{4}$ is $\alpha_s$, and the other one is $\alpha_d$ \cite{Cloude2010}. Using this definition, neither of the surface or double-bounce scatterings has to be selected artificially to be the dominant one \cite{Freeman1998} and the physics of the alpha scattering mechanisms (orthogonality criterion) regulates the decomposition of the ground components. Hence, the respective scattering powers are divided between them accordingly. In the next section, we present the results of applying this estimation approach to all campaign sites of the AirMOSS mission.

\section{Results}
\label{results}
In this section, we present the results of applying the new decomposition approach to observations from all the campaign sites from AirMOSS mission. These include 167 campaigns across all the 10 sites introduced in Table \ref{campaigns}. First, we present statistics across all sites. Then, we discuss example cases from some sites and provide directions for future retrieval algorithm efforts.

\subsection{Overall Statistics}

We use three relative indices to compare the contribution of the scattering mechanisms across different sites. The indices are defined based on the span of the coherency matrix of each scattering component:
\begin{eqnarray}
	\nonumber P_s && = f_s (cos^2\alpha_s + sin^2\alpha_s)\\
	\nonumber P_d && = f_d (cos^2\alpha_d + sin^2\alpha_d)\\
	\nonumber P_v && = f_v\\
	P_T && = \frac{1}{2}(\langle|S_{HH} + S_{VV}|^2\rangle + 2\langle|S_{HV}|^2\rangle + \langle|S_{VV}|^2\rangle)
\end{eqnarray}
in which $P_s$, $P_v$, and $P_d$ represent the span of the coherency matrix for surface, vegetation, and double-bounce scattering, and $P_T$ is the total span of the observation coherency matrix. The three relative indices are $P_s/P_T$, $P_v/P_T$, and $P_d/P_T$ which show the normalized contribution of surface, vegetation and double-bounce scattering, respectively. Figure \ref{rel_hists} shows the marginal probability distribution of these relative indices from one campaign across each of the sites in 2015, 2014, and 2013. The selected campaigns fall around the same time of the year for each site. Each line in Figure \ref{rel_hists} represents data from all the pixels within one campaign for each site. Comparison of the relative contribution probability density functions across the years indicates stable conditions across each site for the same seasons of the years. The interannual variability is small (at least as evident in the marginal probability distributions).

\begin{figure*}
	\centering
	\includegraphics[width=.8\linewidth]{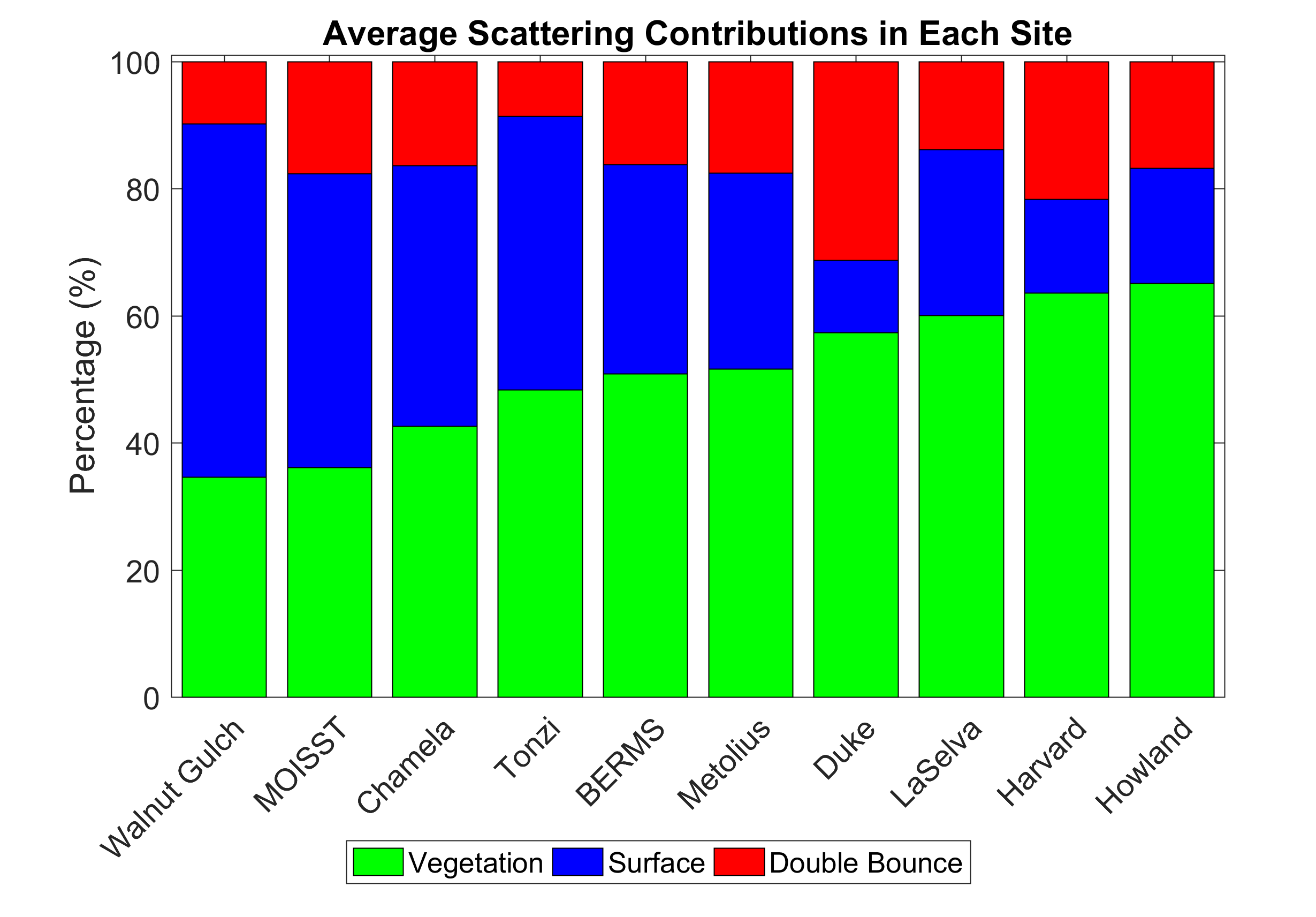}
	\caption{Average contribution of each scattering mechanism toward the total power in each site across all campaigns. Sites are sorted based on vegetation contribution increasing from left to right.}
	\label{BarChartAllCampaigns}
\end{figure*}

Figure \ref{BarChartAllCampaigns} shows the average scattering contribution of each of the three mechanisms in each site across all the campaigns. This figure shows how much of the received power in the SAR instrument is originated from different scattering mechanisms and provides insight for future retrieval algorithm development efforts on where to place the focus in terms of soil moisture or vegetation parameters retrieval. 

Several conclusions can be drawn from these figures:

\begin{itemize}
	\item In dense forest (such as the LaSelva, Harvard, and Howland sites) the vegetation volume scattering is comparatively large and the mode of distribution is between 55-65\% (Figure \ref{rel_hists}, bottom three rows). Meanwhile, the surface contribution is relatively small (mode around 20-30\%). Thus, even though P-band has relatively high penetration through the vegetation and surface compared to commonly used higher frequency ranges, in dense forest  much of the backscatter is still generated by the vegetation and double-bounce scattering (in these cases  10-15\%). 
	
	In the Howland and Harvard sites, which are boreal transitional / mixed forests, the mode of relative vegetation volume scattering is the highest (65-67\%) compared to others, which results in the lowest contribution from the surface in these sites as well (Figure \ref{rel_hists}, bottom two rows leftmost column). 
	
	\item In contrast to dense forest sites, the distribution of the relative vegetation volume scattering contribution is wider in less dense forests such as the Metolius, BERMS, and Duke sites. Although the mode of the magnitude of vegetation volume scattering is relatively high (60\%), the distribution is wider compared to the dense forest sites. Meanwhile, the double-bounce scattering has a wider distribution and a higher mode which shows that the lower density of the forest allows for more interaction of the radar signal between the ground surface and vegetation. Among the three sites Duke has a larger mode for double-bounce. This can be contributed to taller trees across this site compared to Metolius and BERMS. 
	
	\begin{figure}
		\centering
		\includegraphics[width=1\linewidth]{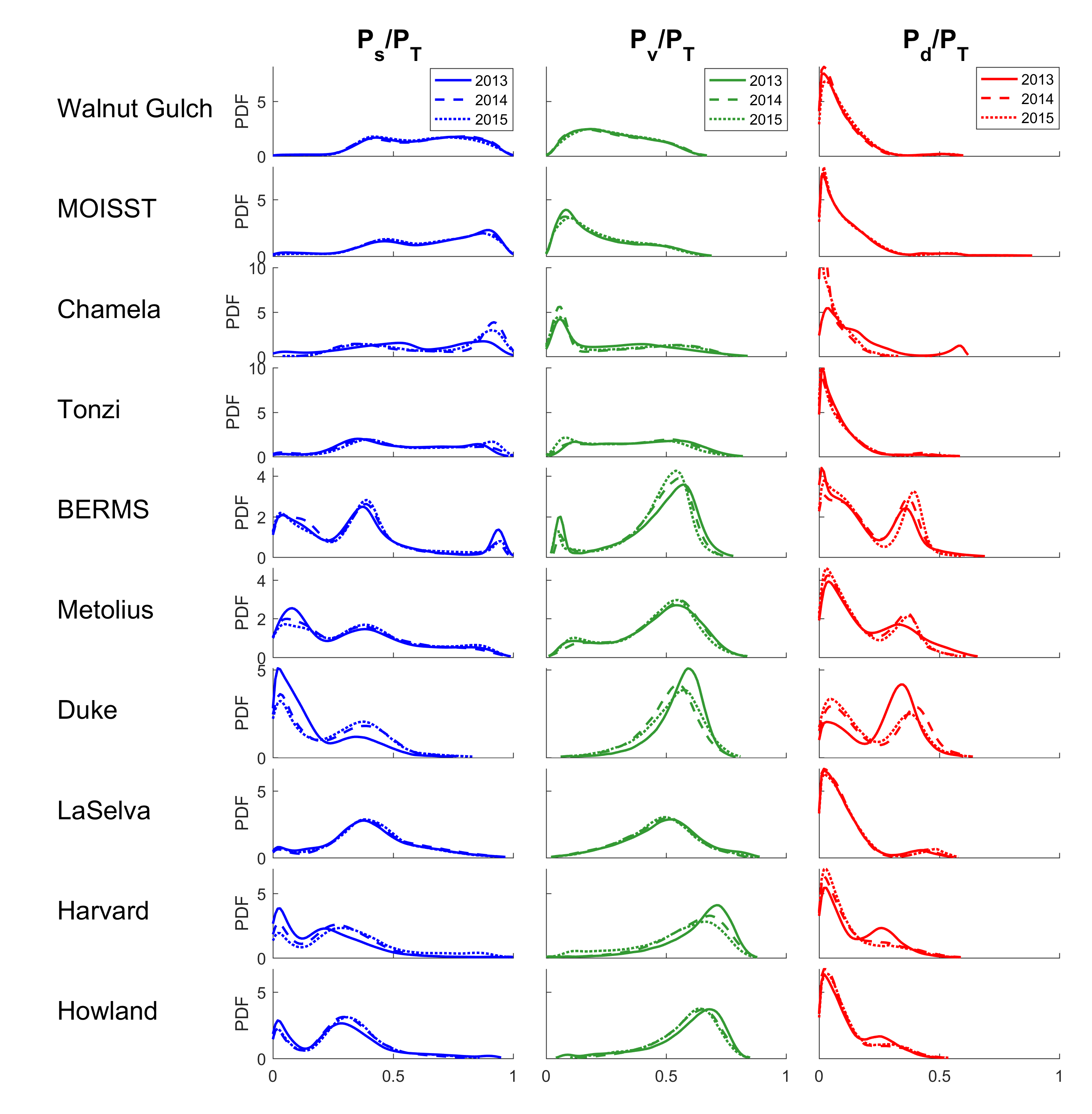}
		\caption{Probability Density Function (PDF) of relative contribution of Surface (left), Vegetation (center), and Double-bounce (right) scattering in the total observed power. Each row represents data from one of the sites, and different lines represent the campaign in different years: solid line shows 2015, dashed line shows 2014, and dotted line shows 2013. Sites are sorted based on the median of the relative vegetation volume scattering, the smallest on the top row.}
		\label{rel_hists}
	\end{figure}

	\item The Tonzi Ranch site has woody savanna land cover. In comparison to Metolius, this site has an even wider distribution of vegetation volume scattering contribution, and a smaller mode. Relative to Metolius, Tonzi shows a much smaller double-bounce scattering (higher values of double-bounce near zero) which is expected given that there is no vegetation cover in some parts of this site. The Chamela site (a dry subtropical forest) has a similar pattern to the Tonzi site. This is likely due to the contributions of cropland and woody savanna pixels over the Chamela flight swath (which results in smaller double-bounce contribution).
	
	\item Lastly, MOISST and Walnut Gulch are the two sites with the least amount of vegetation coverage. MOISST is mostly grassland and cropland while Walnut Gulch is grassland and shrubland. As is evident in this figure, MOISST and Walnut Gulch have relatively wide distribution of surface scattering (similar to Tonzi and Chamela sites) while the mode of surface scattering distribution is larger for Walnut Gulch. Both sites show a relatively wide distribution of vegetation volume scattering with a mode much smaller compared to other sites. 
\end{itemize}

\subsection{Seasonal Scattering Patterns}

In regions with strong seasonal variations of temperature and/or precipitation, vegetation structure and water content can also have significant seasonality. This seasonality changes the interaction of radar signal within the surface - vegetation medium (double-bounce scattering) as well as direct scattering from vegetation and surface. Changes in vegetation structure have a significant implication for retrieving soil moisture or vegetation dielectric properties. Here, we focus on the Tonzi Ranch site, which has strong hydroclimatic seasonality. 

\begin{figure}
	\centering
	\includegraphics[width=.8\linewidth]{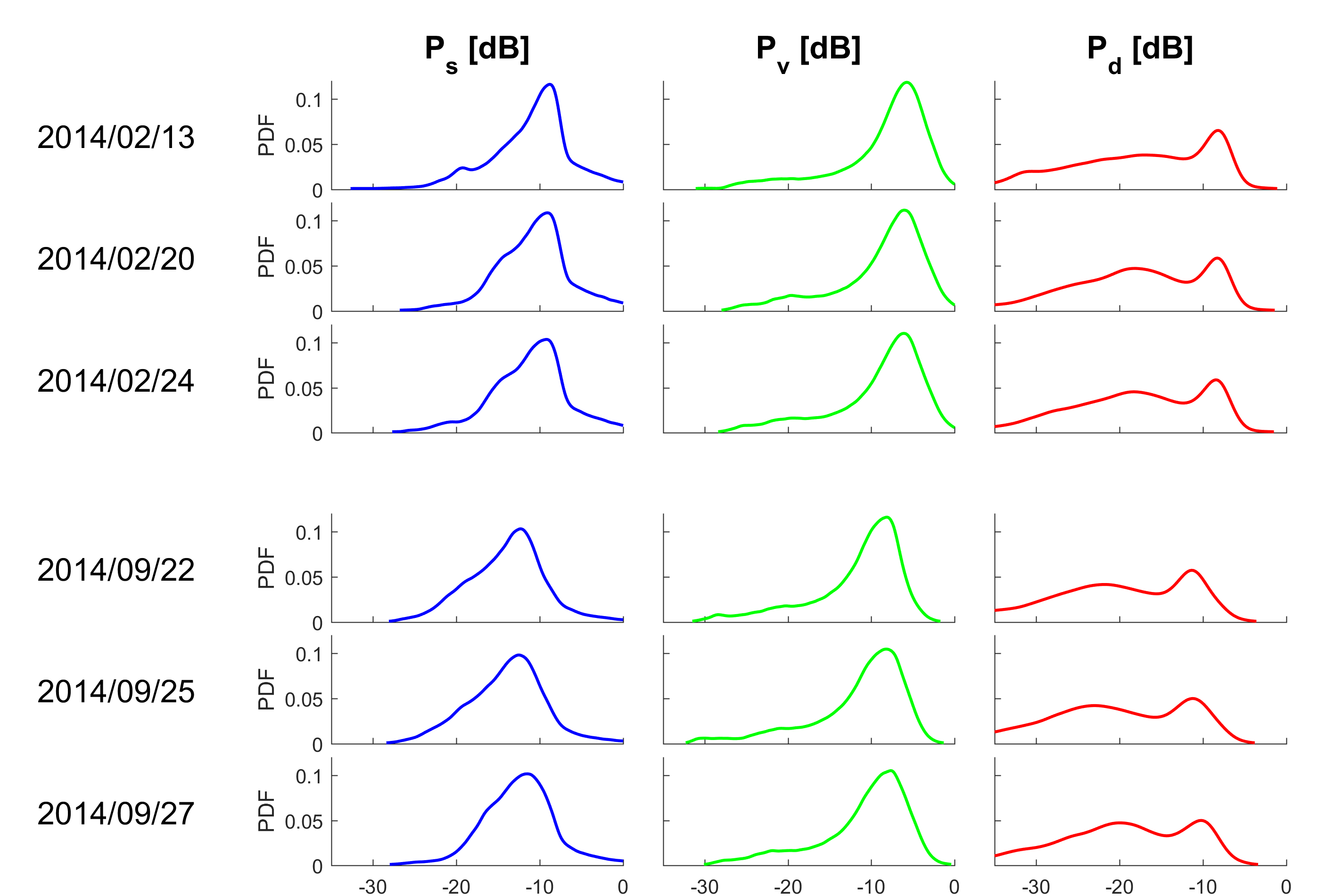}
	\caption{Decomposition results from six flights in two different campaigns (wet and dry season) at Tonzi Ranch. Columns are similar to Figure \ref{rel_hists}, but show the absolute power contribution in [dB]}
	\label{Tonzi_2014_seasonality_dB}
\end{figure}

Tonzi Ranch has a Mediterranean climate with wet winters and dry summers. In 2014, AirMOSS had two campaigns at Tonzi Ranch, one in February and one in September. These two campaigns provide the opportunity to evaluate the impact of climate seasonality on the contributions from each scattering mechanism. Figure \ref{Tonzi_2014_seasonality_dB} shows the decomposition results from 6 flights, 3 in each campaign season, across Tonzi Ranch. This figure shows the absolute value of each of the $P_s$, $P_v$ and $P_d$ to be able to compare the data across two campaigns in terms of absolute contributions. There is a noticeable change in the distribution of each of the scattering mechanisms between the two seasons. 

\begin{figure}
	\centering
	\includegraphics[width=.7\linewidth]{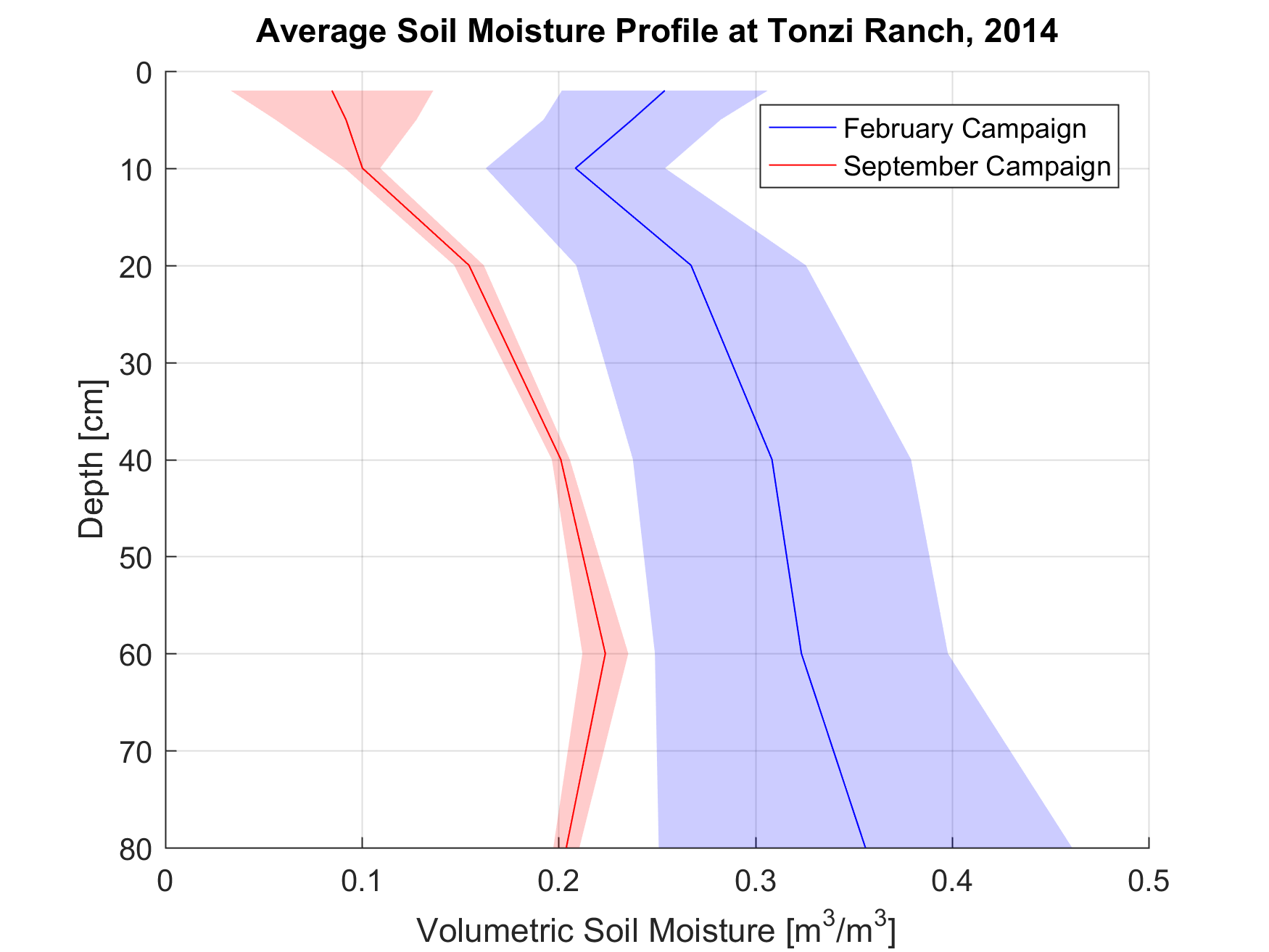}
	\caption{Mean (solid line) and one standard deviation (shading) of soil moisture profile across two campaign seasons at Tonzi Ranch}
	\label{Tonzi_2014_SM}
\end{figure}

The seasonal variations in the scattering terms can be compared to seasonal variations in soil moisture. Measurements from three in-situ soil moisture probes that have been installed across the flight lines of the campaign and provide half-hourly measurements at seven different depths from 2 cm to 80 cm are used here \cite{Romano2013, Cuenca2016}. Data from 10 days prior to the first day of each campaign and during the campaign have been averaged to produce Figure \ref{Tonzi_2014_SM}, which shows the average soil moisture profile during each campaign based on in situ measurements. Soil moisture profiles have significantly changed between the two campaigns (as expected), with drier values occurring in September in the dry season. Moreover, the profile has more variability in the wet season compared to the dry season. 

\begin{figure}
	\centering
	\includegraphics[width=.8\linewidth]{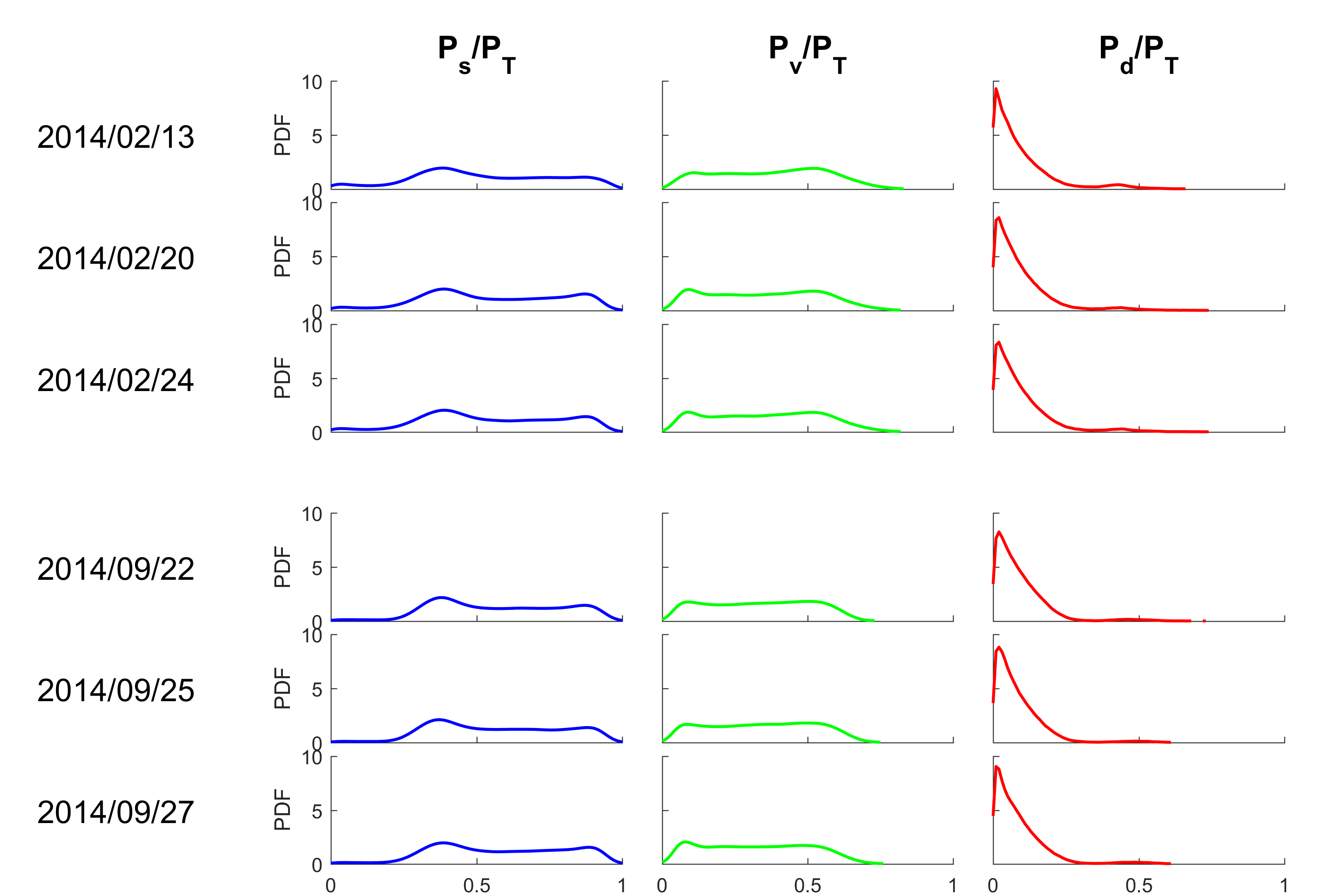}
	\caption{Similar to Figure \ref{Tonzi_2014_seasonality_dB} but showing the relative contributions.}
	\label{Tonzi_2014_seasonality}
\end{figure}

Figure \ref{Tonzi_2014_seasonality_dB} shows that in the September campaign all three scattering mechanisms have decreased. The median of double-bounce, vegetation and surface scattering has decreased 16\%, 25\% and 36\%, respectively. Two main factors contribute to the changes in the relative scattering contributions: vegetation structure and dielectric properties, as well as soil water content. Both factors have played a role. The soil moisture content is lower, and so is the direct surface backscattering. Moreover, the vegetation volume scattering is also lower, since the dry soil results in a drier vegetation cover and disappearance of grasslands which results in less scattering surface. These two contribute to the decrease in the double-bounce scattering which is more pronounced here. However, the relative scattering contributions (shown in Figure \ref{Tonzi_2014_seasonality}) show that the relative contribution of surface scattering ($P_s/P_T$) increased by 10\% in the September campaign. While the absolute scattering has decreased, the surface scattering has more contribution to the total backscattered power. This is consistent with the fact that the decrease in vegetation cover reduces the signal attenuation and enhances signal penetration to the ground. This comparison highlights the importance of a dynamic vegetation volume scattering model that takes into account the seasonality of vegetation structure.

\subsection{Focus Regions}

In this section, we analyze the results from two study regions and discuss the spatial patterns in the decompositions. The goal is to provide insight in the performance of the decomposition approach across different land covers, and inherent limitations that can be improved in the future.

\subsubsection{Case Study 1: Walnut Gulch}

\begin{figure}
	\centering
	\includegraphics[width=.8\linewidth]{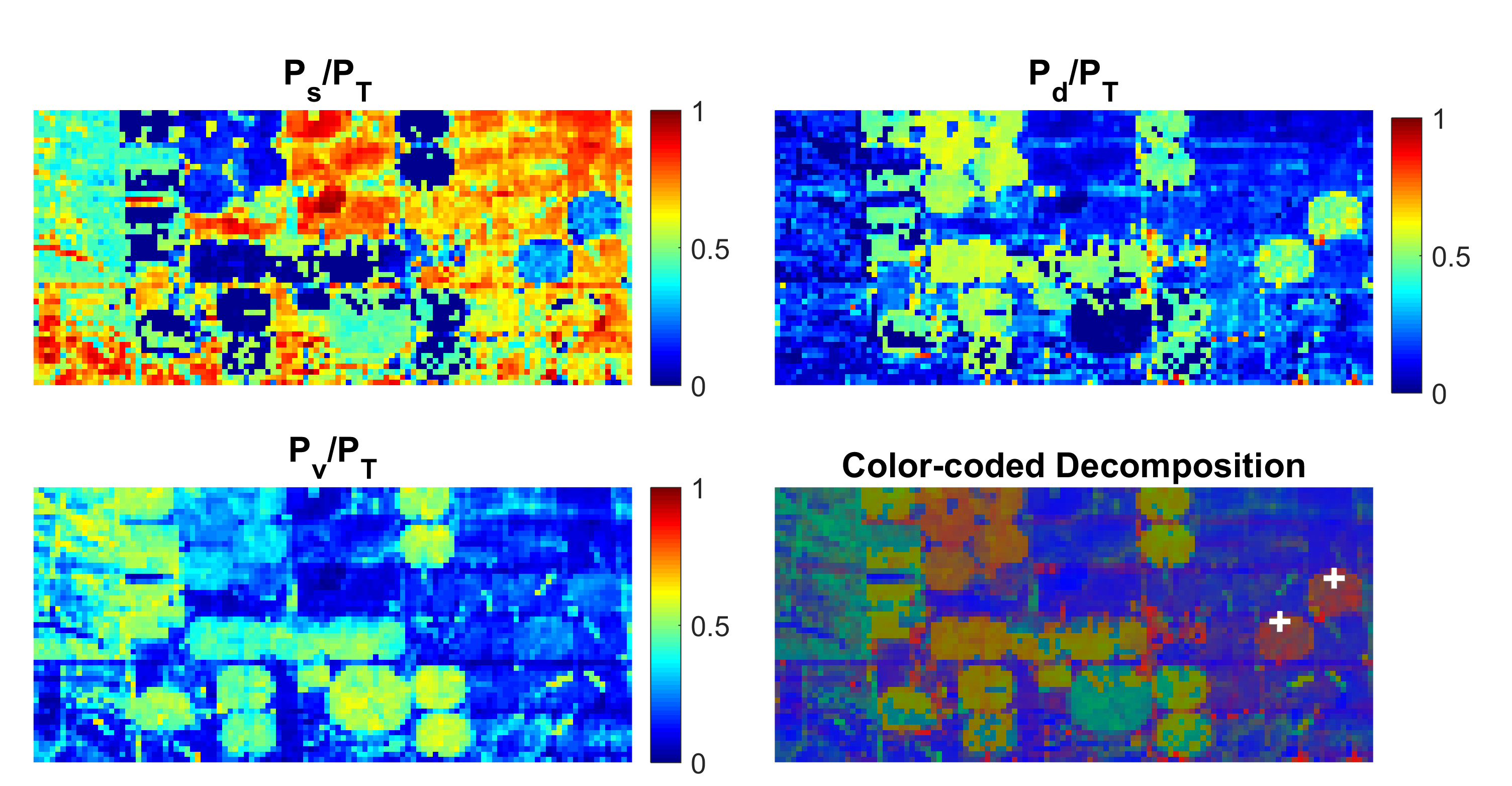}
	\caption{Relative contribution of surface, vegetation and double-bounce scattering mechanisms across the focus region in Walnut Gulch. Lower left panel shows the RGB-coded decomposition contributions (R: double-bounce, G: vegetation, B: surface). White pixels are either missing or filtered because of the range of incidence angle. The white crosses on the RGB panel show two pivotal irrigation fields that are analyzed in the text.}
	\label{Walnut}
\end{figure}

This case shows a focus region of approximately 4.6 km x 8.6 km located near Walnut Gulch in the southeastern part of the state of Arizona in US. The land cover of this region is mostly bare land and shrubs with several pivotal agricultural fields located in this domain. We use observation from a flight campaign conducted on Aug 8, 2015. Figure \ref{Walnut} shows the results of the decomposition approach in terms of the relative contribution of surface, double-bounce and volume scattering. This figure also presents an RGB-coded decomposition map which summarizes the contribution of each scattering mechanism in a single image (red: double-bounce scattering, green: volume scattering, and blue: surface scattering). 

The circular patterns visible in each of the panels show different agricultural fields that use pivotal irrigation systems. In bare land pixels, the surface contribution is large and vegetation and double-bounce are almost zero. In vegetated pixels, however, vegetation volume scattering is high, double-bounce scattering is noticeable, and surface scattering is very small.

However, there are two pivotal systems to the right of the domain (denoted by white + signs on the bottom right panel of Figure \ref{Walnut}) that have very small vegetation volume scattering and relatively large double-bounce scattering together with small surface scattering. This pattern is consistent between the three flight days in a week time period of this campaign (only one day is shown here). Recognizing that vegetation volume scattering contribution is not zero, but is low compared to other active pivotal farms, it is possible that vegetation in those pixels has a structure that is not consistent with the randomly-oriented model; therefore, scattering power is shifted to double-bounce. This can happen for example in vertically aligned stacks of a cereal crop that leads to high double bounce contribution but low random volume scattering \cite{ Jagdhuber2016}. Analysis of the phase of the $S_{HH}S_{VV}^*$ across these two fields show that the phase ranges between -14$^\circ$ to -25$^\circ$ for one of the fields, and between -32$^\circ$ and -55$^\circ$ for the second one. The significant departure of the phase from zero shows that the pixels in these two fields are not bare soil, and support the argument that vegetation coverage is present and causing high double-bounce scattering, but without a random volume-like scattering contribution.

\subsubsection{Case Study 2: Tonzi Ranch}

\begin{figure}
	\centering
	\includegraphics[width=.8\linewidth]{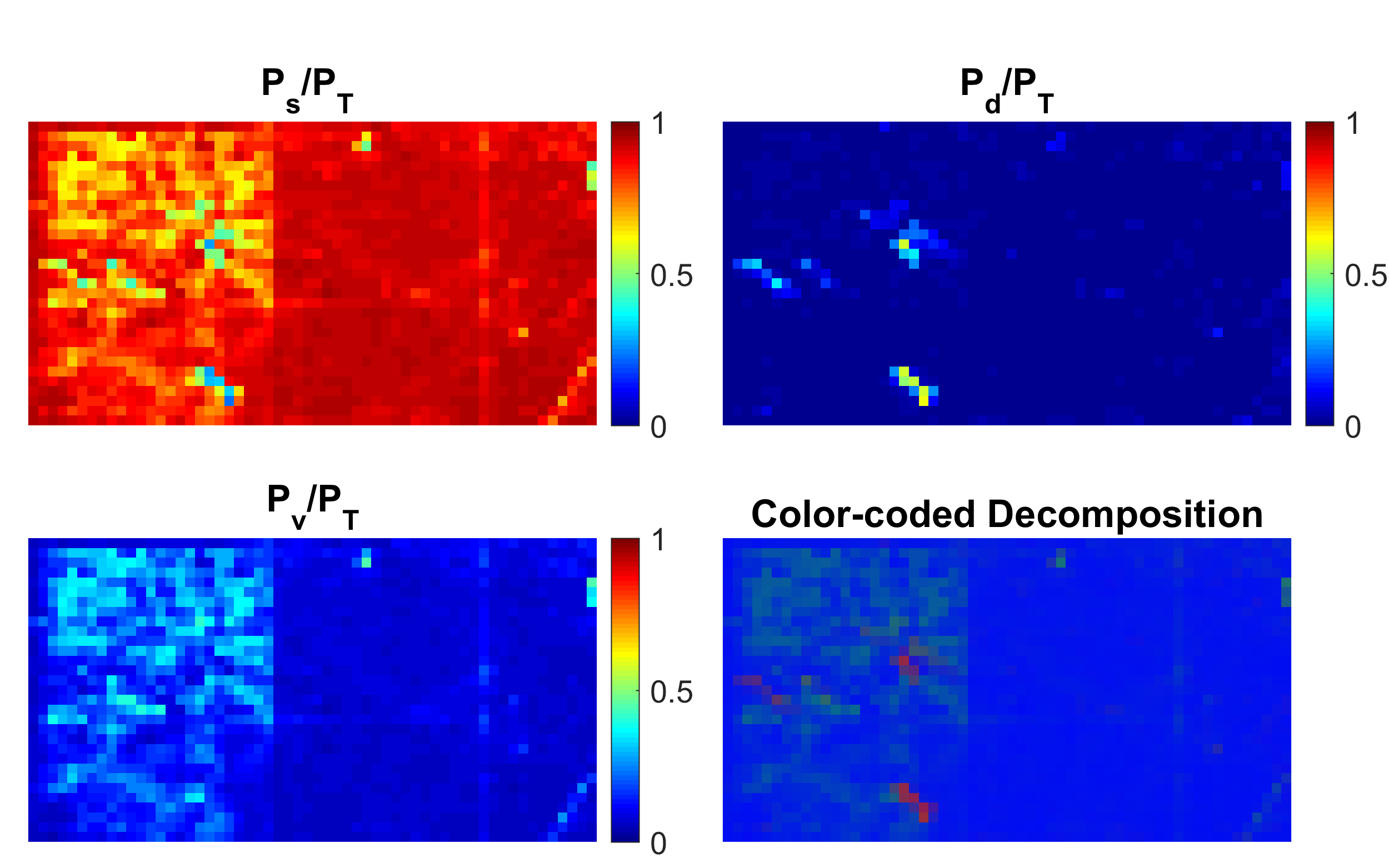}
	\caption{Similar to Figure \ref{Walnut} but for Focus region in Tonzi Ranch.}
	\label{Tonzi}
\end{figure}

The second focus region considered is located in Tonzi Ranch in the central part of the state of California. It covers an area of about 2.8 km x 4.2 km. Observations for this region are from Feb 10, 2015. Figure \ref{Tonzi} shows the result of the decomposition in this domain, similar to Figure \ref{Walnut} for the first focus region. This domain is also dominated by bare land and low vegetated pixels that show up clearly in the decomposition. However, in the left half of the domain the decomposition pattern appears random, and changes dramatically between adjacent pixels. Analyzing the vegetation volume scattering contribution does not provide any clue on why two adjacent regions can have such different surface scattering contributions. 

\begin{figure}
	\centering
	\includegraphics[width=.8\linewidth]{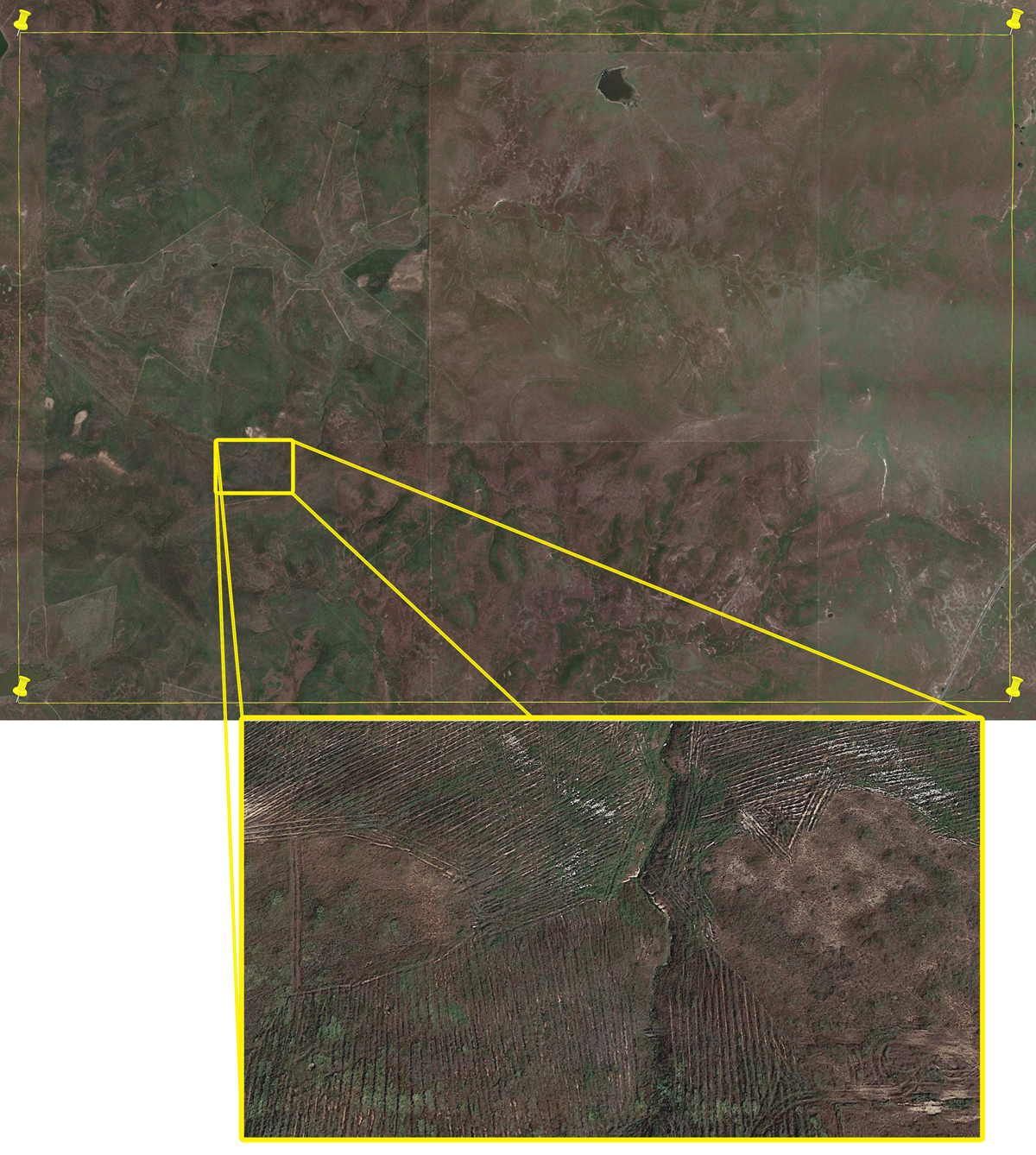}
	\caption{Satellite-based imagery of the Tonzi Ranch study box from Google Earth. The yellow bounding box with pins at the corner shows the boundary of the study box in Figure \ref{Tonzi}. The inset box shows a smaller domain with finer details.}
	\label{Tonzi_box_Google}
\end{figure}

Figure \ref{Tonzi_box_Google} shows satellite-based imagery from the same focus region based on Google Earth. This image is captured within a 6 week time window from the time of P-band observations. Also shown is a close-up zoom of a small but representative part of the domain. Complex patterns of tillage are present in the left side of the domain, while there is no similar pattern on the right side. These patterns are likely contributing to the random pattern of surface scattering in the decomposition results. Surface roughness can also contribute to high cross-polarization observation, and this is visible in the $P_V/P_T$ estimates in Figure \ref{Tonzi}. These signatures of surface roughness show that assuming constant surface roughness for retrievals from P-band observations even across small domains can result in biased estimations of surface and vegetation contribution and lead to erroneous soil moisture retrievals. 

\section{Discussions and Conclusion}
\label{conclusion}
In this study, we apply a hybrid (combined model- and eigen- based) decomposition technique to P-band SAR observations. We estimate the contributions of different scattering mechanisms across a wide range of biomes. The methodology builds on the previously published three component decomposition approach to combine two techniques (model-based with eigen-based decomposition) and uses the coherent fully polarimetric observations. This approach is applied to polarimetric observations from AirMOSS mission, which are phase-calibrated using a data-driven calibration scheme.

The hybrid approach does not make allometric assumptions to estimate vegetation volume scattering. While a similar approach has been previously applied to L-band observations across agricultural fields to estimate surface soil moisture \cite{Jagdhuber2015}, our study is the first of its kind to explore the application of a hybrid decomposition approach with P-band observations across a wide range of vegetation covers, in particular, dense forests. These decompositions enable future development of retrieval algorithms to estimate soil moisture profiles in the presence of vegetation canopy exploiting fully polarimetric observations. 

Applying the estimation approach to observations from 167 AirMOSS campaign flights across 10 sites with diverse vegetation types, we estimated the scattering contributions of surface, vegetation, and double-bounce mechanisms. The relative scattering contributions of these three mechanisms characterize different biomes at each site consistent with expectations from in-situ observations. Observations at Walnut Gulch showing the effects of pivotal irrigation system support the validity of our decomposition. The median of the vegetation volume scattering contribution ranged from 32\% at Walnut Gulch to 67\% at Howland, while surface scattering median ranged from 15\% to 55\%. The double-bounce contributions were more constant across sites, with an average median of 13\% (except for Duke site with a mode of ~30\%).

We also investigate the results at specific site/campaigns and report the advantages and limitations of applying this decomposition framework. The difference in seasonal patterns of the estimations show the difference in vegetation volume scattering contributions across different seasons. Moreover, the relative contributions of surface and double-bounce scattering in different biomes provide valuable insight for designing retrieval algorithms for soil moisture profile using low-frequency polarimetric observations. Such a decomposition can help to better characterize the inverse problem of soil moisture profile estimation with fewer unknowns using one of the surface or double-bounce scattering contributions or both. 

The vegetation model can be improved to include a second dipole distribution for vegetation volume scattering and retrieving their relative weights. The second distribution can be targeted at tall forest trees that have a different interaction with low-frequency microwave signals compared to leaves and branches, and have been shown to have a higher double-bounce contribution \cite{Moghaddam1995, Lucas2004}. 

Surface roughness is known to be more important in the observations at low-frequencies compared to higher ones. However, estimation of surface roughness is still a challenge. As we show using a case study at Tonzi Ranch, at P-band the effect of surface roughness may be significant and it is necessary to take into account a spatially variable surface roughness across the domain.

\section*{Acknowledgment}

Authors wish to thank members of the AirMOSS science team for their inputs and constructive feedbacks on different parts of this research. Funding for this study is provided by a NASA grant to Massachusetts Institute of Technology as part of the AirMOSS mission.

\bibliographystyle{IEEETran}
\bibliography{AirMOSSBib}

\end{document}